\documentclass[prl, twocolumn, superscriptaddress, nofootinbib, nobibnotes]{revtex4-2}

\usepackage[normalem]{ulem}

\usepackage[english]{babel}
\usepackage{hyperref}

\usepackage[pdftex]{graphicx}
\usepackage{color}

\usepackage{amsmath}
\usepackage{amssymb}
\usepackage{mathtools}
\usepackage{braket}
\usepackage{stmaryrd}

\usepackage[sort&compress]{natbib}

\newcommand{\un}[1]{\ensuremath{\,\mathrm{#1}}}
\renewcommand{\v}[1]{\ensuremath{\boldsymbol{#1}}}

\newcommand{\fig}[1]{Figure~\ref{fig:#1}}

\newcommand{\lr}[1]{\ensuremath{\left( #1 \right)}}

\newcommand{\I}{\mathrm{i}}

\newcommand{\mc}{\mathcal}

\newcommand{\affil}{\affiliation{Instituto de Ciencias F\'isicas, Universidad Nacional Aut\'onoma de M\'exico, Cuernavaca, México}}

\begin{document}

\title{Edge-state transport in gapped bilayer graphene}

\author{Jesús~Arturo~Sánchez-Sánchez}
\email{jasanchez@icf.unam.mx}
\affil

\author{Thomas~Stegmann}
\email{stegmann@icf.unam.mx}
\affil

\date{\today}

\begin{abstract}
We investigate electronic transport in gapped bilayer graphene (gBLG) devices. For certain edge terminations -- typically a combination of zigzag, armchair, and bearded types -- we observe edge state conduction within the band gap, which is opened by a potential bias between the two layers. The edge states can generate a non-local resistance, in line with recent experiments \cite{inspo}. Band structure calculations of gBLG nanoribbons corroborate the existence of the edge states, whose edge localization can be switched by tuning the electron energy. Their existence strongly depends on the edge termination and does not originate from a topological bulk-boundary correspondence.
\end{abstract}

\maketitle

\section{Introduction}

After entering the stage about two decades ago \cite{1}, graphene quickly gained notoriety in condensed matter physics and material science \cite{fame1,fame2,fame3}. Early on, graphene was seen as a possible alternative to silicon-based electronic components that continuously are getting smaller. However, a problem is that pristine graphene can not be electrically turned off due to the lack of a band gap \cite{wallace1947band,1}. Hence, several mechanisms have been proposed throughout the years to open an energy gap in graphene-based systems. Such mechanisms include periodic distortions of the lattice that break chiral symmetry \cite{gamayun2018valley,kekO}, breaking the sublattice symmetry of the honeycomb lattice \cite{symm-break1,sublattice-break}, long-range perturbations of the graphene layer\cite{long-range-gap}, and applied potential bias \cite{bias1,bias2}. 

In particular, the application of a potential bias in bilayer graphene (BLG) opens a tunable energy gap by breaking the inversion symmetry of the graphene layers \cite{bias3} creating gapped BLG (gBLG). BLG results from stacking two graphene monolayers which creates one of the simplest yet interesting van der Waals structures. BLG has a semi-metallic behavior with a parabolic low-energy dispersion \cite{2,3,4} in contrast to the linear dispersion of monolayer graphene \cite{wallace1947band}. This parabolic behavior occurs in the most common and stable phase known as AB or Bernal stacking, which will be used in this article and will be referred to simply as BLG. This material exhibits both excellent electrical and thermal conductivity at room temperature \cite{5,6}, mechanical stiffness, strength and flexibility \cite{7,8}, high transparency \cite{9} and allows the tuning of its electrical properties via doping and external gating \cite{10,11,12,13,naumisstrained}. 

The 'simplicity' of the potential bias method makes gBLG a fairly accessible system and a very attractive research subject. Case in point, the origin of non-local transport measurements in gBLG devices is debated. Recently, non-local transport measurements were performed in dual-gated gBLG devices before and after etching of the edges \cite{inspo}. After etching, the non-local resistance exceeds  Ohmic contributions and is attributed to the appearance of conducting edge channels. The observed edge dependence of non-local transport in these gBLG devices differs from the previous explanations centered around the valley Hall effect \cite{valleyhall1,valleyhall2}.

\begin{figure}[h]
   \centering
   \includegraphics[width=\linewidth]{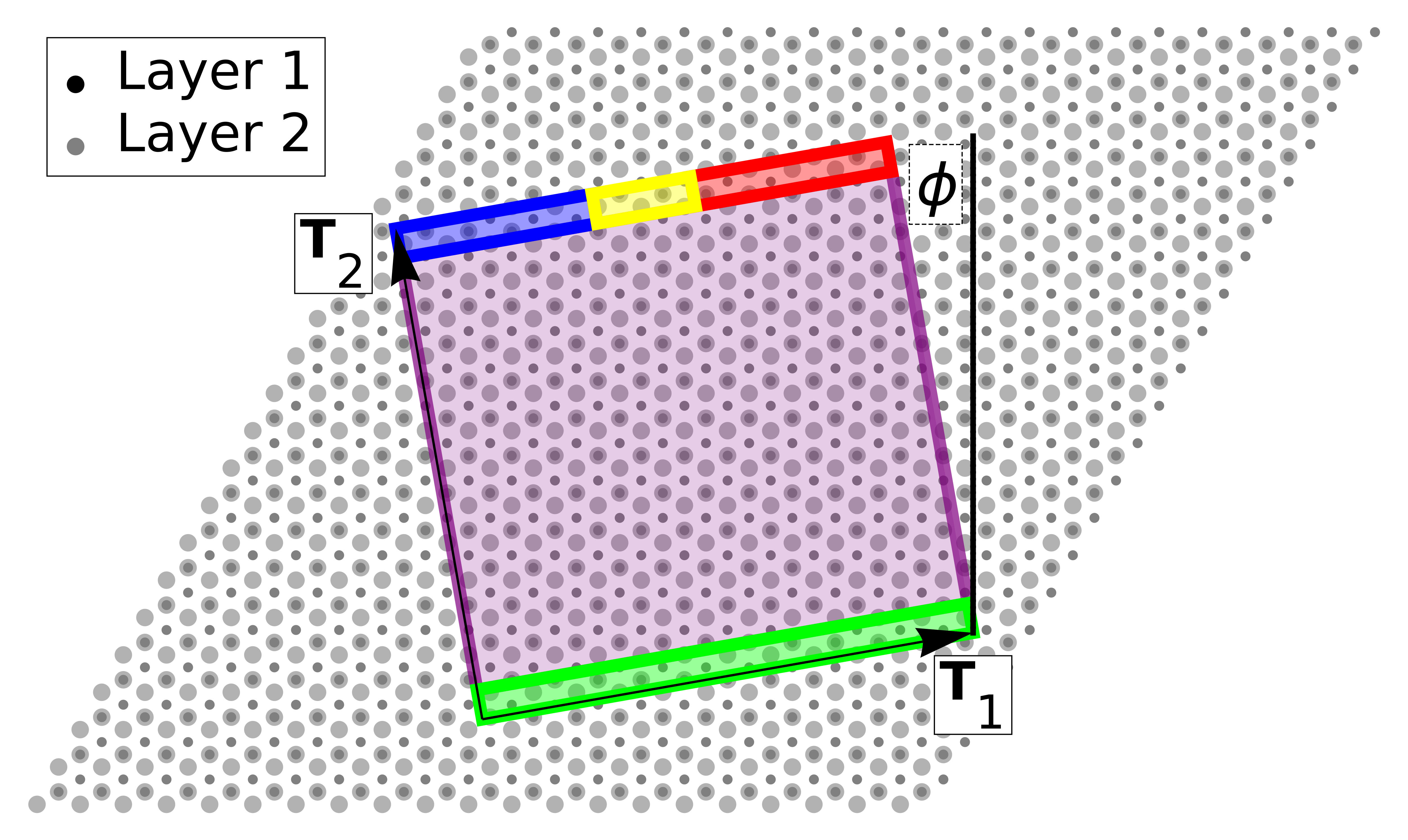}
\caption{Squematic of the gBLG devices. A 'cookie cutter', defined by the vectors $\mathbf{T}_1$ and $\mathbf{T}_2$, is rotated by the angle $\phi$ above an AB stacked BLG lattice. The colored regions correspond to the source contact $S$ (green), and three drain contacts $D_L$ (blue), $D_M$ (yellow) and $D_R$(red).}
   \label{fig:device}
\end{figure}

In this article, we study edge transport in gBLG devices, sketched in \fig{device}, in the vicinity of the energy gap produced by an applied potential bias between the graphene layers. We model gBLG devices with different edge terminations to achieve a more general description of their transport properties and their dependence on the edge termination. Transport calculations show electronic transport inside the energy gap due to edge states. These edge states have a bidirectional nature and their existence depends crucially on the edge termination of the devices. Some edge terminations support edge states while others do not, leaving the system gapped. The edge dependence of these states contrasts with the valley Hall effect, which originates from bulk properties. With our transport calculations we are able to explain the experimental non-local transport measurements reported recently in gBLG \cite{inspo}. Finally, we support this explanation with band structure calculations in gBLG nanoribbons with various edge terminations.
 
\section{System \& Methods}

\subsection{gBLG  devices}
 
The process of constructing the gBLG devices is shown in Figure \ref{fig:device}. We begin with a graphene monolayer (black circles) with lattice vectors $\mathbf{a}_1=a_0 ( \sqrt{3},0)$ and $\mathbf{a}_2=a_0(\frac{\sqrt{3}}{2},\frac{3}{2})$, where $a_0=0.142\un{nm}$ is the carbon-carbon distance. A second graphene monolayer (gray circles) is shifted by $\frac{1}{3}(\mathbf{a}_1+\mathbf{a}_2)$ and placed $d_{0}=0.335\un{nm}$ above the first monolayer in order to get an AB stacking. This BLG lattice serves as a template for a 'cookie cutter' (purple area) spanned by the vectors $\mathbf{T}_1=L_1(\cos{\phi},\sin{\phi})$ and $\mathbf{T}_2=L_2(-\sin{\phi},\cos{\phi})$. In this work, we focus mainly on gBLG devices with a size of $L_1=200a_0=28.4 \un{nm}$ and $L_2=250a_0=35.5\un{nm}$. The 'cookie cutter' is rotated by the angle $\phi=[-1^{\circ},31^{\circ}]$ in order to produce gBLG devices with different edge terminations. The case of $\phi=30^{\circ}$ corresponds to a zigzag edge along $\mathbf{T}_2$, although some contributions of bearded edges are present, because the value of $L_1$ is not a multiple of the size of a carbon hexagon. As shown in Figure~\ref{fig:device}, a source contact $S$ (green area) located at one edge injects electrons into the device, which are collected by three drain contacts on the opposite edge; a left drain $D_L$ (blue area), a middle drain $D_M$ (yellow area) and a right drain $D_R$ (red area). This contact setup enables us to characterize the electronic transport along $\mathbf{T}_2$.

\subsection{Tight-Binding model}

In order to model gBLG systems, we use the tight-binding Hamiltonian
\begin{equation}
    \label{H}
    H= \sum_{n} \epsilon_{n} \ket{n} \bra{n} +\sum_{n,m}  t_{nm} \ket{n} \bra{m} + \text{H.c.} 
\end{equation}
where $\ket{n}$ indicates the atomic state localized on the carbon atom at $\v{r}_n$. The first term of the Hamiltonian corresponds to the onsite energies $\epsilon_n$ and the second term takes into account the couplings between the carbon atoms, which are given by the Slater-Koster formula \cite{slater} 
\begin{equation}
    \label{t_nm}
    t_{nm}= \cos^{2} (\gamma) V_{pp\sigma}(r_{nm})+ [1-\cos^{2}(\gamma)]V_{pp\pi}(r_{nm}),
\end{equation}
where $\cos\gamma=z_{nm}/r_{nm}$ is the direction cosine of $\v{r}_{nm}=\v{r}_m-\v{r}_n$ along the $z$ direction. The Slater-Koster parameters are defined as 
\begin{equation}
    \label{Slater-Koster}
    \begin{split}
    V_{pp\sigma}(r_{nm})&=V_{pp\sigma}^{0}\exp\left[q_{\sigma} \Bigl (1-\frac{r_{nm}}{d_0} \Bigr)\right],\\
    V_{pp\pi}(r_{nm})&=V_{pp\pi}^{0}\exp\left[q_{\pi} \Bigl(1-\frac{r_{nm}}{a_0} \Bigr) \right]
    \end{split}
\end{equation}
with the tight-binding parameters \cite{param} 
\begin{equation}
\label{TB_parameters}
    \begin{split}
    V_{pp\sigma}^{0}=0.48 \un{eV}, \quad V_{pp\pi}^{0}=-2.7 \un{eV},\\[1mm]
    \frac{q_{\pi}}{a_0}=\frac{q_{\sigma}}{d_0}=22.18 \un{nm}^{-1}. \quad
    \end{split}
\end{equation}
The intralayer couplings are included up to first nearest neighbors, while the interlayer couplings are taken into account up to a cut-off radius of $ 0.614 \un{nm}$.

The onsite energies, $\epsilon_n=\pm \epsilon$, have a constant value in each layer but change sign between the layers in order to take into account the effect of the electrostatic potential generated by gate contacts. This opens a energy gap \cite{2,mccann2007low}
\begin{equation}
    \label{gap}
   \epsilon_g=\frac{2\epsilon V_{pp\sigma}^{0}}{\sqrt{(V_{pp\sigma}^{0})^2+4\epsilon^2}}
\end{equation}
in the electronic structure spanning over the energy range $[-\frac{\epsilon_g}{2},\frac{\epsilon_g}{2}]$. For the rest of this article we will use a value of $\epsilon=0.075  |V_{pp\pi}^{0}|=203 \un{meV}$, which opens an energy gap of $\epsilon_g=310 \un{meV}$. Let it be noted that Equation~\eqref{gap} is derived taking into account only nearest neighbor intralayer couplings and interlayer couplings of vertically aligned sites. In our calculations, we take into account more distant interactions, decreasing the value of the energy gap slightly.

\begin{figure*}[t]
   \centering
   \includegraphics[width=1.0\linewidth]{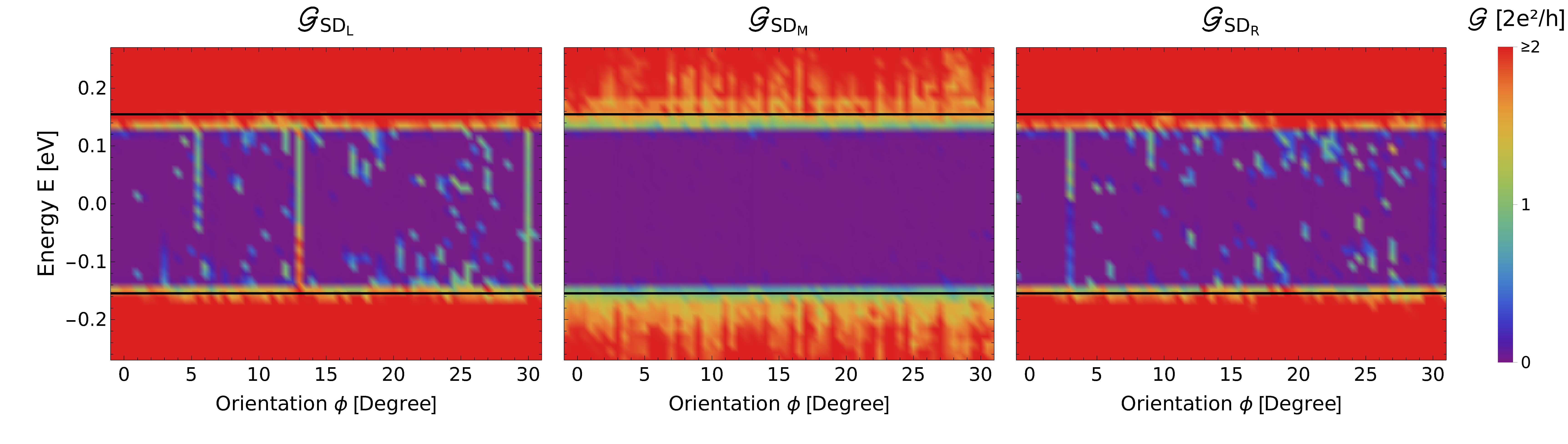}
\caption{Conductance $\mathcal{G}$ between the source $S$ and three drain contacts $D_{L/M/R}$ in gBLG devices as a function of the edge orientation $\phi$ and electron energy $E$. The energy gap is easily identified in $\mathcal{G}_{SD_M}$ because electronic transport is suppressed inside the energy range $[-\frac{\epsilon_g}{2},\frac{\epsilon_g}{2}]$ (black horizontal lines). The transport signals within the gap, which are present in both, $\mathcal{G}_{SD_L}$ and $\mathcal{G}_{SD_R}$, but absent in $\mathcal{G}_{SD_M}$, suggest edge transport.}
 \label{fig:ts}
\end{figure*}

\subsection{Green's function method}

The Green's function method is applied to study the electrical transport in the gBLG devices and here, we briefly summarize the essential equations \cite{negf1,negf2,negf3}. The Green's function of the system is given by
\begin{equation}
    \label{GF}
    G=\Bigl(E-H-\sum_p\Sigma_p\Bigr)^{-1}
\end{equation}
where $E$ is the energy of the injected electrons and $H$ is the tight-binding Hamiltonian \eqref{H}. The contacts are modeled by the so-called wide-band model implying for these a constant energy-independent surface density of states. It is taken into account in the Green's function by means of the self-energies $ \Sigma_{p}= -\I V_{pp\pi}^{0} \sum_{n \in C_{p}} \ket{n}\bra{n}$, where the sum runs over the carbon atoms $C_p$ encompassed by the contact $p=\{S,D_L,D_M,D_R\}$.

The transmission between a pair of contacts, $i$ and $j$, is given by
\begin{equation}
    \label{Transmission}
	T_{ij}(E)= \mathrm{Tr}[\Gamma_{i}G\Gamma_{j}G^{\dagger}],
\end{equation}
where the inscattering function associated to contact $i$ is defined as $\Gamma_{i}= -2\mathrm{Im}(\Sigma_i)$. The transmission determines the conductance for electrons of energy $E$ between the selected pair of contacts, $\mathcal{G}_{ij}= (2e^2/h) \, T_{ij}$. In addition, the local current flowing between the atoms $n$ and $m$ can be calculated with \cite{curr1,curr2}
\begin{equation}
    \label{I}
I_{nm}(E)= \frac{e}{h}\mathrm{Im}\bigl[t^{*}_{nm} (G\Gamma_{S}G^{\dagger})_{nm} \bigr].
\end{equation}

Finally, for the non-local resistance calculations we consider 4 contacts in the device, where electrons are injected by the S contact, collected in the D contact and the voltage between a pair of contacts A and B is calculated. The non-local resistance is calculated as follows
\begin{align}
    \label{Rnl}
R_{\text{NL}}&=\frac{V_{AB}}{I_{SD}}=\frac{h}{2e^2} \frac{\sum_{j} T_{Sj} \lr{\mc{M}_{jA}-\mc{M}_{j B}}}{T_{SD} + \sum_{ij} T_{Di} \mc{M}_{ij} T_{jS}}
\end{align}
where
\begin{equation}
    \label{Mmat}
    \mc{M}^{-1}=
   \begin{cases}
   -T_{ij} \quad &\text{if} \quad i\neq j,\\
   \sum_{k\neq i} T_{ik} \quad &\text{if} \quad i=j
   \end{cases}
\end{equation}
The summation in Equation~\eqref{Rnl} runs only over the contact $A$ and $B$, while the summation in Equation~\eqref{Mmat} includes all contacts. For more details on the derivation see our previous work \cite{nosotros}.

\section{Electronic transport}

\subsection{Conductance}

\begin{figure*}[t]
   \centering
   \includegraphics[width=\linewidth]{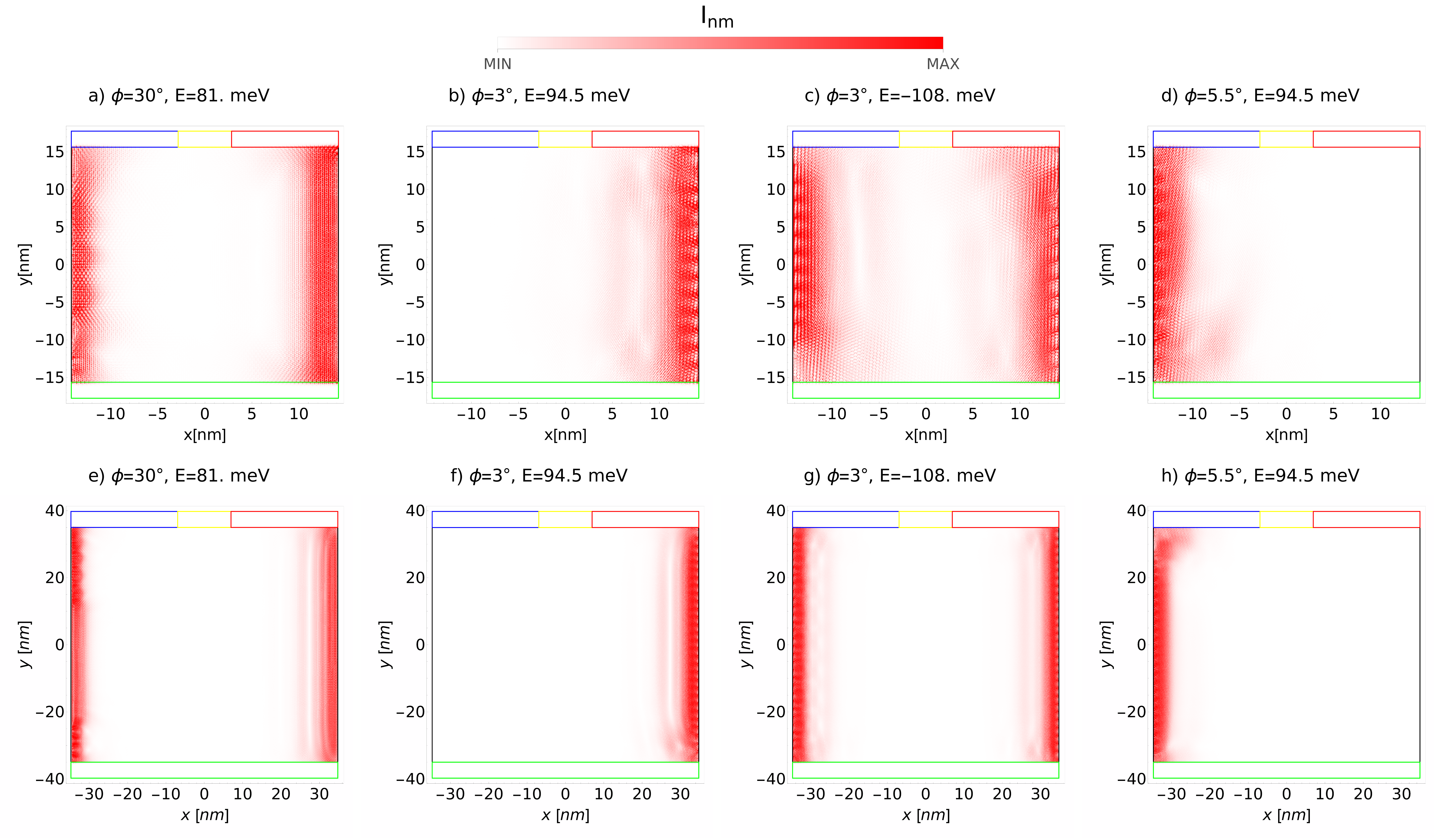}
    \caption{Current density, indicated by the red color shading, for electrons with energy $E$ in gBLG devices with different edge orientations $\phi$. All panels prove that transport in the gap takes place at the edges of the devices. Panels (a-d) are for devices with size $28.4 \times 35.5\un{nm}$, which are used also for the calculations of the conductance and non-local resistance. In panels (e-h) the size of the device is increased to $69.6 \times 79.5 \un{nm}$, but the parameters are otherwise the same to demonstrate that the edge states are rather size independent. The current flow can be localized at both edges (a,c,e,g) or only at a single one (b,d,f,h). Importantly, this edge localization of the current can even be switched by the electron energy (b-c,g-f).}
   \label{fig:I}
\end{figure*}

We begin our study of electronic transport in gBLG devices with Figure~\ref{fig:ts} showing the conductance between the source and drain contacts (their position is sketched in Figure~\ref{fig:device}) as a function of both, energy of the injected electrons $E$ and edge orientation $\phi$. 

First, transport is heavily suppressed inside the gap, which is especially easy to spot in the Figure for $\mathcal{G}_{SD_M}$, because the conductance is negligible for all orientations. This means that the electrons injected at $S$ into the gBLG devices do not reach the middle contact. The size of the transport gap agrees well with its expected energy range $[-\frac{\epsilon_g}{2},\frac{\epsilon_g}{2}]$ according to Equation~\eqref{gap}, indicated by black horizontal lines. Note also that conductance values larger than twice the conductance quantum, appearing rapidly outside the gap, are painted in red, as we focus on transport features inside the gap. Our conductance calculations show also that the electronic transport is bidirectional, $\mathcal{G}_{i\rightarrow j} = \mathcal{G}_{j\rightarrow i}$, which is expected because time reversal symmetry is preserved in the system. Therefore, we refer to the conductance indistinctly as $\mathcal{G}_{ij}$.

The conductance between the source and the lateral contacts, $\mathcal{G}_{SD_L}$ and $\mathcal{G}_{SD_R}$, shows clearly non-zero conductance signatures inside the gap for certain electron energies and edge orientations. Electrons injected at $S$ arrive exclusively at $D_{R/L}$, suggesting edge-dominated transport. If electrons could flow freely across the device, one would expect to find similar results in all drain contacts. Point in case, for energies above and below the energy gap the conductance is essentially the same for all drain contacts, where the main difference is the size of $D_M$ compared to $D_{L,R}$. Notably, some conductance signals extend over the entire energy gap. Non-zero conductance values at $\phi=30^{\circ}$ are expected since the edges have (mostly) zigzag shape that host edge-localized states in graphene and BLG nanoribbons. However, one would not expect in advance that orientations such as $\phi=3^{\circ}$ and $\phi=13^{\circ}$ could have transport across the entire energy gap. Likewise, $\phi=5.5^{\circ}$ shows non-zero conductance values for more than half of the energy gap.

So far, our conductance calculations shown in Figure~\ref{fig:ts} confirm a strong dependence on the edge termination of the electronic transport in gBLG devices. Also, we can begin to understand the asymmetry between $\mathcal{G}_{SD_L}$ and  $\mathcal{G}_{SD_R}$ as a product of the asymmetry of the edges themselves, since the width of the device is not a multiple of the size of the carbon hexagons. 

\subsection{Local current flow}

\begin{figure*}[t]
   \centering
   \includegraphics[width=0.9\linewidth]{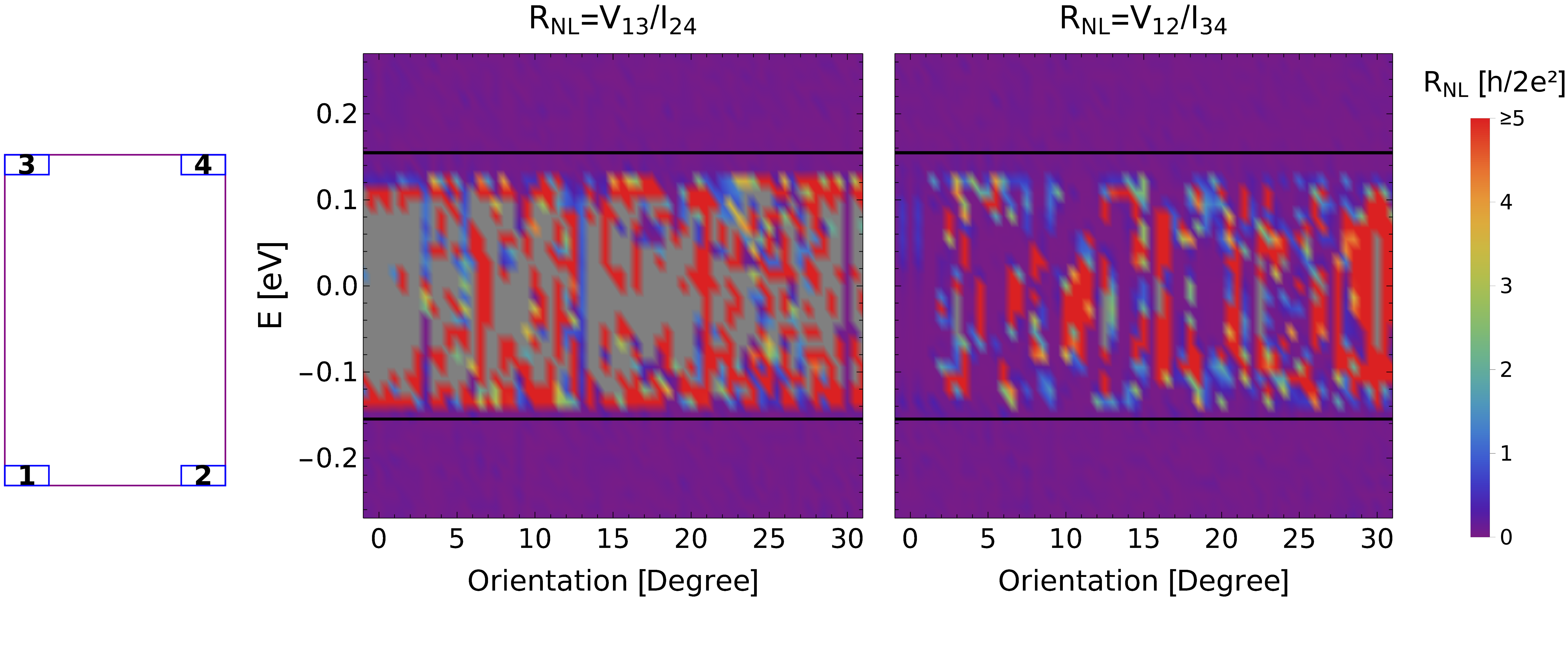}
    \caption{Non-local resistance $R_{\text{NL}}$ as a function of edge orientation $\phi$ and energy $E$. A setup with 4 contacts at the corners of the device is used and two distinct contact wirings are studied. Clear signals of $R_\text{NL}$ are found within the band gap due to edge transport. Exceedingly large values of $R_{NL}$ (gray regions) are due to a negligible current between $S$ and $D$.}
   \label{fig:RNLs}
\end{figure*}

In order to confirm that the conductance to the lateral drain contacts is the product of edge transport, we perform current flow calculations via Equation~\eqref{I} for a selection of gBLG devices. We identify each system by their edge orientation $\phi$ and specify the energy of the injected electrons $E$. Figure \ref{fig:I} shows the calculated current flow for gBLG devices with different edge orientations; $\phi=30^{\circ}$ in panels (a,e), which corresponds (mostly) to a zigzag edge, $\phi=3^{\circ}$ in (b-c,f-g) and $\phi=5.5^{\circ}$ in (d,h). Systems in panels (a-d) have the same size ($28.4 \times 35.5\un{nm}$) as those used for the calculations of the conductance and non-local resistance, while in (e-h) the size of the devices is increased to  $69.6 \times 79.5 \un{nm}$.

Our hypothesis of edge conduction is proven correct as the current flow patterns show that electrons injected at $S$ (green area) flow exclusively at the edges of the gBLG devices and are collected at the lateral drain contacts (blue and red areas). Panels (a,c,e,g) show that both edges are able to support current flow simultaneously, while in (b,f) it is only at the right edge and in (d,h) only at the left one. Panels (b-c,f-g) show that depending on the electron energy, the current can flow on both or on a single edge. This demonstrates that for a given edge termination one can tune the conductance by changing the energy of the injected electrons. 

Comparing the current flow patterns in panels (a-d) with those shown in panels (e-h), where only the size of the devices has been increased while the edge orientation is identical, proves that the edge states are rather size independent. Importantly, we have evaluated also the corresponding conductance values in the larger devices and found them essentially the same as in the smaller devices. All current flow patterns are in agreement with the conductance values reported in Figure \ref{fig:ts} and confirm that current flow takes place exclusively at the edges of the gBLG devices for energies inside the energy gap.

\subsection{Non-local resistance}

In order to study further the transport properties of the gBLG devices, we analyze in Figure~\ref{fig:RNLs} the non-local resistance $R_\text{NL}$, calculated by means of Equation~\eqref{Rnl}. This enables us to link our findings to experiments such as those reported in \cite{inspo} for etched gBLG devices with a Hall bar geometry. In contrast to the previous sections, here we use a device with contacts at the corners, allowing for transport along all edges. Any of those four contacts can serve as source $S$, drain $D$ or a voltage probe $A$ and $B$, although we focus on two wirings as indicated in Figure \ref{fig:RNLs}. In both cases, we find clear signals of the non-local resistance inside the band gap. This indicates that transport is present at some edges of the device, depending on the parameters $E$ and $\phi$. Note that similar to the coloring of Figure~\ref{fig:ts}, $R_{NL}$ values larger than 5 times the conductance quantum are colored red. Exceedingly large values of the non-local resistance, $R_{NL} \geq 100  \frac{h}{2e^2}$, are indicated in gray color and can be explained by negligible values of the denominator $I_{SD}$ in  Equation~\eqref{Rnl}, meaning a current flow is not possible due to absence of edge states for the given system parameters. 

We have also evaluated the cross conductions, $\mathcal{G}_{14}$ and $\mathcal{G}_{23}$, and found that they are zero, indicating again that the transport is not through the bulk but along the edges. We emphasize that the edge states are bidirectional and can be localized only at some of the device edges. Our findings give a possible explanation of the non-local resistance observed in gBLG Hall bars after etching \cite{inspo}, because the etching process modifies the edges and can induce edge states.
 
\section{Band structure calculations}\label{bands}

\subsection{gBLG  nanoribbons}

\begin{figure}[h]
   \centering
   \includegraphics[width=0.9\linewidth]{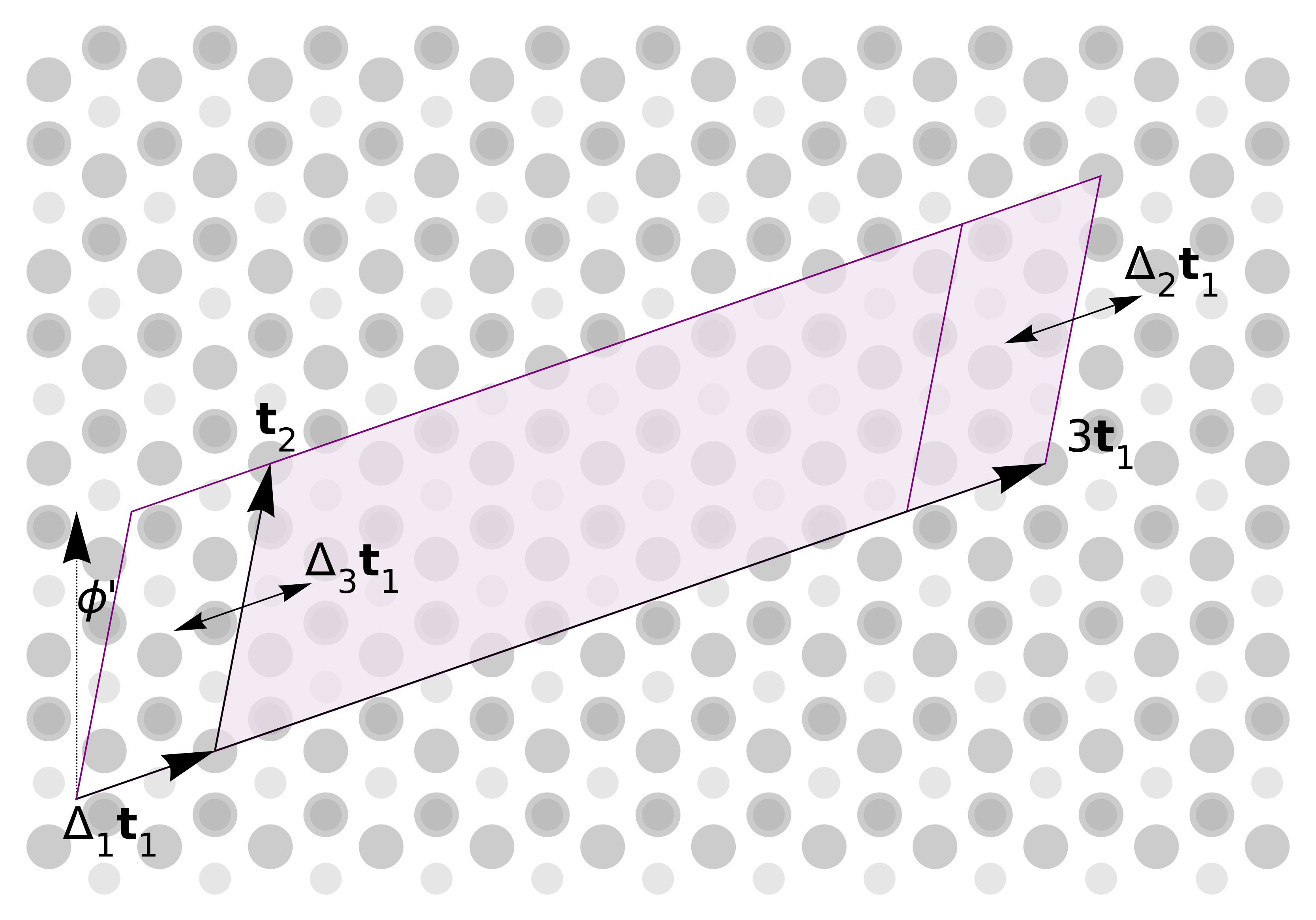}
\caption{gBLG nanoribbon construction. Superlattice vectors $\mathbf{t}_1$ and $\mathbf{t}_2$ define the unit cell of the nanoribbon. The edge of the nanoribbon is along the periodic direction $\mathbf{t}_2$ and the finite width is $3\mathbf{t}_1$. The angle $\phi'$ links a given nanoribbon to a device with edge orientation $\phi$. The edges of the nanoribbon are modified by the parameters $\Delta_1, \Delta_2$ and $\Delta_3$.}
   \label{fig:ribbon}
\end{figure}

\begin{figure*}[t]
   \centering
   \includegraphics[width=\linewidth]{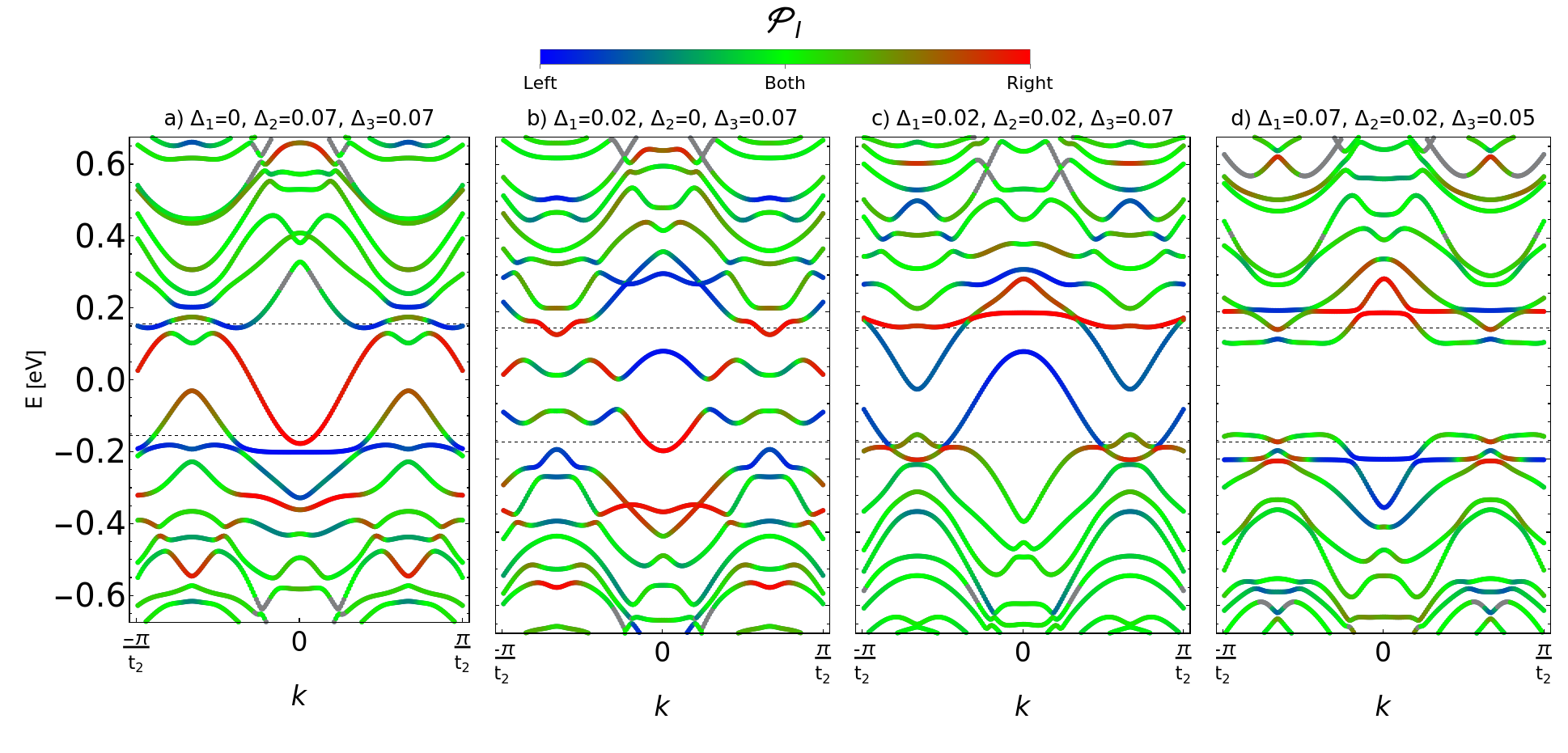}
\caption{Energy bands for the $13^{\circ}$-gBLG nanoribbon (generated by $p=7$ and $q=3$, leading to $\phi' = 13.0039^{\circ}$) and varying edge terminations through the parameters $\Delta_{1/2/3}$. The coloring of the bands correspond to the weighted inverse participation ratio $\mathcal{P}_I$, which indicates if the states are localized on a single (blue/red) or both (green) edges. Gray colors are used for states that are delocalized (distributed over $>45\%$ of the nanoribbon's atoms). The parameters in panel (c) can explain the conductance in the corresponding device: transport along the left edge due to the existence of edge localized states.}
   \label{fig:bands13}
\end{figure*}

To substantiate the transport results, we perform band structure calculations of gBLG nanoribbons. Using the originally constructed BLG lattice (Figure \ref{fig:device}), we create nanoribbons with a finite width of 3 unit cells in one direction and periodicity in the other direction. The edges along this periodic direction are the ones that will approximate the edges of the gBLG devices in the 3-drain configuration. 

We create the unit cell of the nanoribbon (purple area) as depicted in Figure \ref{fig:ribbon}. With a pair of positive integers $p$ and $q$ we identify different gBLG nanoribbons with superlattice vectors $\mathbf{t}_1=p\mathbf{a}_1+q\mathbf{a}_2$ and $\mathbf{t}_2=-q\mathbf{a}_1+(p+q)\mathbf{a}_2$. Periodic conditions are imposed in the direction of $\mathbf{t}_2$ and a finite width of $3\mathbf{t}_1$ is used. The edge orientation of a gBLG nanoribbon is given by the angle
\begin{equation}
    \label{angle}
\cos\phi'=\frac{\mathbf{t}_2 \cdot \hat{y}}{ |\mathbf{t}_2|}=\frac{\sqrt{3}(p+q)}{2\sqrt{p^2 + pq+q^2}}.
\end{equation}
The values of $p$ and $q$ will be chosen such that 
the edge orientation is almost the same as in the finite gBLG devices, $\phi\approx\phi'$. In other words, we approximate $\mathbf{T}_2$ with $\mathbf{t}_2$, because for the band structure calculations periodic nanoribbons are needed, a restriction that is not necessary in the case of finite devices.

In addition, the gBLG nanoribbons are constructed with small width variations in order to study the edge dependence of their electronic structure. The width variation is controlled by three parameters $\Delta_1$, $\Delta_2$ and $\Delta_3$, as shown in Figure \ref{fig:ribbon}. The entire unit cell can be moved by $\Delta_1 \mathbf{t}_1$, changing both edges simultaneously. We can also change the edges independently from one another as the right edge can be modified by $\Delta_2 \mathbf{t}_1$, while the left edge can be modified by $\Delta_3\mathbf{t}_1$.

\subsection{Band structure of gBLG nanorribons}

\begin{figure*}[t]
   \centering
   \includegraphics[width=\linewidth]{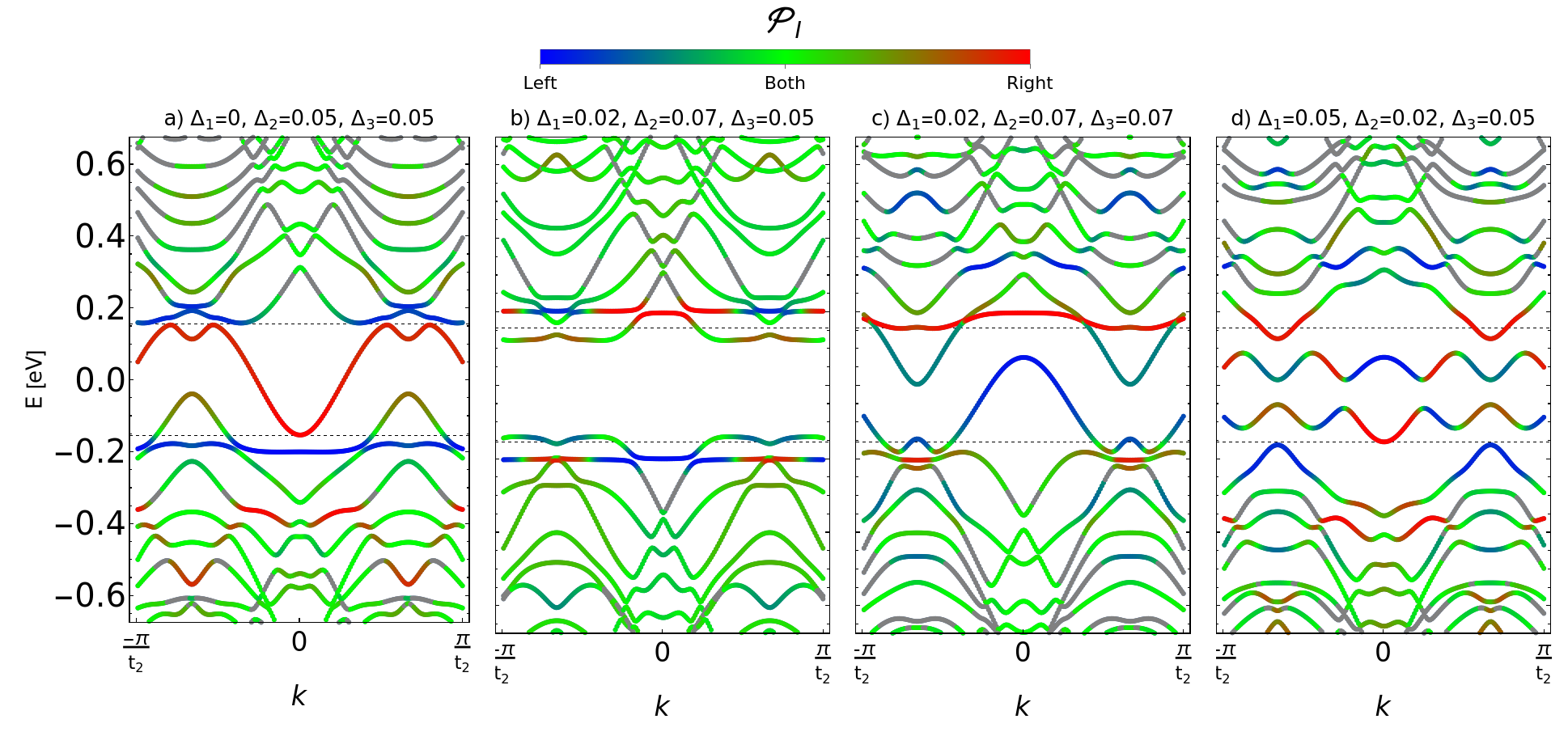}
\caption{Energy bands for the $3^{\circ}$-gBLG nanoribbon (generated by $p=6$ and $q=5$, leading to $\phi'=3.00449^{\circ}$) and varying edge terminations. The coloring of the bands corresponds to the weighted inverse participation ratio $\mathcal{P}_I$. The edge termination in (a) explains the conductance in the band gap at $\phi=3^{\circ}$: transport along both edges for negative energies and transport along the right edge for positive energies due to the existence of edge localized states.}
   \label{fig:bands3}
\end{figure*}

The band structures of the gBLG nanoribbons with edge orientations of $13^\circ$ and $3^\circ$ are shown in Figures \ref{fig:bands13} and \ref{fig:bands3}, respectively. The color of the bands corresponds to the weighted inverse participation ratio $\mathcal{P}_I$, which indicates whether an eigenstate is localized or not (as the usual inverse participation ratio \cite{ipr}), but also if it is localized on a single or on both edges. To achieve this, for each eigenstate at a given $k$-point we calculate
\begin{equation}
    \label{PI}
\mathcal{P}_I=\frac{\sum_{i=1}^{N}x_i|\psi_i|^4}{\sum_{i=1}^{N}|\psi_i|^4},
\end{equation}
where $N$ is the number of atoms in the nanoribbon and the vector $(x_1,x_2,...,x_N)$ encodes how close is each atom to the edges of the nanoribbon. Additionally, we explored the parameter space created by $\Delta_{1/2/3}$ and report the most significant band structures.

The band structure of the $13^\circ$-gBLG nanoribbon (approximated by $p=7$ and $q=3$, leading to $\phi'=13.0039^{\circ}$) is shown in Figure \ref{fig:bands13} for various values of $\Delta_{1/2/3}$. The band structure is strongly dependent on the edge termination, particularly notable for the energy states inside the band gap. However, panel (c) can explain the conductance results of the $\phi=13^{\circ}$ gBLG device: current flowing only on the left edge inside the energy gap due to the existence of energy states localized at the left edge. To continue our analysis, we show in Figure \ref{fig:bands3} the band structure of the $3^\circ$-gBLG nanoribbon (approximated by $p=6$ and $q=5$, leading to $\phi'=3.00449^{\circ}$). In panel (a), we find inside the gap at positive energies a state that is localized only at the right edge, while for negative energies it is localized at both edges. This explains the localization of the current flow reported in Figure \ref{fig:I} (b-c,f-g). For other values of $\Delta_{1/2/3}$ the edge localized states can disappear, or shift to other edges, highlighting that the reported states depend on slight changes of the edge termination. 

Note that the band structures in Figures \ref{fig:bands13} and \ref{fig:bands3} show similar features, although the edge orientations,  $13^\circ$ and $3^\circ$, are quite different. Both gBLG nanoribbons can host a gapped spectrum (Figure \ref{fig:bands13} (d) and Figure \ref{fig:bands3} (b)) as well as edge localized states that does not cover the entire energy gap (Figure \ref{fig:bands13} (b) and Figure \ref{fig:bands3} (d)). The most striking result is that the band structures that support the transport properties for one edge orientation can almost be reproduced exactly by another edge orientation. For example, Figure \ref{fig:bands13} (c) explains the transport at $\phi= 13^\circ$ but Figure \ref{fig:bands3} (c) is almost identical. The same situation applies to Figure \ref{fig:bands13} (a) and Figure \ref{fig:bands3} (a) in the context of the transport at $\phi=3^{\circ}$. This pattern lets us conclude that only a finite number of edge states exist in the gBLG system, but these states are available for most edge orientations. This observation also implicates that details in the Figures \ref{fig:ts} and \ref{fig:RNLs} may change for a different construction of the gBLG devices, but the overall message -- edge state transport within the gap -- will remain the same. We have analyzed the different edge terminations that host edge states and found that they are not purely zigzag, armchair, or bearded, but rather a mixture of all three.

\section{Conclusions}

In this paper, we studied the transport phenomena inside the energy gap of gBLG devices with a combination of a tight-binding Hamiltonian and the Green's function method, and support our findings with band structure calculations. The gBLG devices were created in a way that produces different edge terminations via the parameter $\phi$ (Figure~\ref{fig:device}). Conductance signatures were found inside the band gap (Figure~\ref{fig:ts}) and the local current flow proved that these are due to edge transport (Figure~\ref{fig:I}). 

We selected two specific edge orientations, $\phi= 13^\circ$ and $\phi= 3^\circ$, and performed band structure calculations to demonstrate that indeed edge-localized states appear inside the band gap (Figures~\ref{fig:bands13} and \ref{fig:bands3}). However, these states turned out to be rather fragile because small changes of the edge termination through the parameters $\Delta_{1/2/3}$ can make them appear or disappear. Importantly, the energy bands found at $13^\circ$ could also be observed at $3^\circ$, suggesting that only a finite number of edge states is available to the device. 

The edge states are bi-directional and lead to a non-local resistance (Figure~\ref{fig:RNLs}), because they can be localized only at certain edges of the device. A non-local resistance is measured in gBLG dvices after etching \cite{inspo}, which we explain by the fact that etching affects strongly the edge termination. We also note that edge transport was reported by us previously in twisted bilayer graphene \cite{nosotros2}. Although band gaps are opened through different mechanisms, both systems exhibit conducting edge states that strongly depend on the edge termination. The found edge states cannot be attributed to topological features of gBLG (as also pointed out by the authors of \cite{inspo}) because they depend sensitively on the edge termination. For technological applications, it may be advantageous that the edge localization can be switched by the electron energy (Figure~\ref{fig:I} (b,c)), which in principle allows to control the current flow through electrostatic gating. 

\section{Acknowledgments}

JAS-S gratefully acknowledges a SECIHTI graduate scholarship. We gratefully acknowledge funding from UNAM-PAPIIT IG10072 and the 'Fundación Marcos Moshinsky'.

\end{document}